\documentstyle[12pt,a4]{article}
\newcommand{\be}{\begin{equation}}
\newcommand{\ee}{\end{equation}}
\newcommand{\bea}{\begin{eqnarray}}
\newcommand{\eea}{\end{eqnarray}}
\newcommand{\bed}{\begin{displaymath}}
\newcommand{\eed}{\end{displaymath}}

\newcommand{\bc}{\begin{center}}
\newcommand{\ec}{\end{center}}

\newcommand{\br}{({\bf r})}
\newcommand{\brp}{({\bf r'})}
\begin{document}
\bc
\vspace*{-3.2cm}
\begin{verbatim}
To appear in:
"Density Functional Methods in Chemistry"
(ACS series, 1996)
\end{verbatim}\vspace{1.2cm}
{\LARGE \bf Conventional Quantum Chemical Correlation Energy
versus Density-Functional Correlation Energy\\}
\vspace{1.6cm}
E.K.U. Gross, M. Petersilka and T. Grabo\\
\vspace{1cm}
Institut f\"ur Theoretische Physik, Universit\"at W\"urzburg, \\
Am Hubland, D-97074 W\"urzburg, Germany \\
\vspace{2.1cm}
\ec
\hfill
\parbox{12cm}{
We analyze the difference between the correlation energy as defined
within the conventional quantum chemistry framework and its namesake in
density-functional theory. Both quantities are rigorously defined concepts;
one finds that $E_c^{QC} \geq E_c^{DFT}$. We give numerical and analytical
arguments suggesting that the numerical difference between the two rigorous
quantities is small. Finally, approximate density functional correlation
energies resulting from some popular correlation energy functionals are
compared with the conventional quantum chemistry values.}
\hfill
\vspace{2cm}

{\Large \bf Inroduction}

In quantum chemistry (QC), the exact correlation energy is
traditionally defined as the difference between the exact total energy
and the total selfconsistent Hartree-Fock (HF) energy:
\be \label{QCdefinition}
E_{c, exact}^{QC} := E_{tot, exact} - E_{tot}^{HF} \quad .
\ee
Within the framework of density-functional theory
(DFT)~\cite{DrGr90,PaYa89}, on the other
hand, the correlation energy is a functional of the density
$E_c^{DFT}\left[ \rho \right]$.
The exact DFT correlation energy is then obtained by inserting the
exact ground-state density of the system considered into the
functional $E_c^{DFT}\left[ \rho \right]$, i.~e.~
\be \label{EcexactDFT}
E_{c, exact}^{DFT} = E_c^{DFT} \left[ \rho_{exact} \right] \quad .
\ee
In practice, of course, neither the quantum chemical correlation
energy (\ref{QCdefinition}) nor the DFT correlation energy
(\ref{EcexactDFT}) are known exactly. Nevertheless, both quantities are
rigorously defined concepts.

The aim of the following section is to give a coherent overwiev of how
the correlation energy is defined in the DFT literature~[3--13]
and how this quantity is related to the conventional QC correlation energy.
It will turn out that $E_{c, exact}^{QC}$ and
$E_{c, exact}^{DFT}$ are generally not identical and that $E_{c,
exact}^{QC} \geq E_{c, exact}^{DFT}$. Furthermore we will give an
analytical argument indicating that
the difference between the two exact quantities is very small.

In the last section we compare the numerical values of {\em
approximate} conventional QC correlation energies with {\em
approximate} DFT correlation energies resulting from some popular DFT
correlation energy functionals. It turns out that the difference
between DFT correlation energies and QC correlation energies is
smallest for the correlation energy functional of Colle and
Salvetti~\cite{coll&sall75,coll&sall79}
further indicating~\cite{grab&gros95} that the results
obtained with this functional are closest to the exact ones.

\vspace{16pt}{\Large \bf Basic Formalism}

We are concerned with Coulomb systems described by the Hamiltonian
\be \label{Hamiltonian}
\hat{H} = \hat{T} + \hat{W}_{Clb} +\hat{V}
\ee
where (atomic units are used throughout)
\bea
 \hat{T} &=& \sum_{i=1}^{N} \left(-\frac{1}{2} {\bf\nabla}_i^{2} \right) \\
 \hat{W}_{Clb} &=& \frac{1}{2} \sum_{\stackrel{i,j=1}{i\neq j}}^{N}
             \frac{1}{\left| {\bf r}_i - {\bf r}_j \right|}  \\
 \hat{V} &=& \sum_{i=1}^{N} v \left( {\bf r}_j \right) \quad .
\eea
To keep the following derivation as simple as possible, we choose to
work with the traditional Hohenberg-Kohn~\cite{HoKo64} formulation
rather than the constrained-search representation~\cite{Le79, Li82,
Le82} of DFT. In particular, all ground-state wavefunctions
(interacting as well as non-interacting) are assumed to be non-degenerate.
By virtue of the Hohenberg-Kohn theorem~\cite{HoKo64} the ground-state
density
$\rho$ uniquely determines the external potential $v=v\left[ \rho
\right]$ and the ground-state wave function $\Psi \left[ \rho
\right]$.
If $v_{0} \left({\bf r}\right)$ is a given external potential
characterizing a particular physical system, the Hohenberg-Kohn
total-energy functional is defined as
\be \label{HK}
E_{v_0}\left[ \rho \right] =
\langle \Psi [\rho]|\,\hat{T}+\hat{W}_{Clb}+\hat{V}_0 \,|\Psi [\rho]
\rangle \quad .
\ee
As an immediate consequence of the Rayleigh-Ritz principle, the
total-energy functional (\ref{HK}) is minimized by the exact
ground-state density $\rho_{exact}$ corresponding to the potential
$v_0$, the minimum value being the exact ground-state energy, i.~e.~
\be \label{Etot,exact}
E_{tot, exact} = E_{v_0} \left[ \rho_{exact} \right] \quad .
\ee
In the context of the Kohn-Sham (KS) scheme~\cite{KoSh65} the
total-energy functional is usually written as
\be \label{1}
 E_{v_0}[\rho] = T_{s}[\rho] + \int \! \rho({\bf r}) v_0({\bf r}) \, d^3r
+ \frac{1}{2} \int\! \int \! \frac{\rho({\bf r}) \rho({\bf r}')}
                           {|{\bf r - r'}|}     \, d^3r \, d^3r'
+ E_{xc}[\rho]
\ee
where $T_s\left[ \rho \right]$ is the kinetic-energy functional of
non-interacting particles.
By virtue of the Hohenberg-Kohn theorem, applied to non-interacting systems,
the density $\rho$ uniquely determines the single-particle potential
$v_s \left[ \rho \right]$ and the ground-state Slater-determinant
\be \label{6}
\Phi^{KS}[\rho] = \frac{1}{\sqrt{N!}} \det \left\{ \varphi_{j \sigma}^{KS}
[\rho] \right\}
\ee
and hence $T_s \left[ \rho \right]$ is given by
\bea \label{Ts}
T_{s}[\rho] &=& \langle \Phi^{KS}[\rho]|\hat{T}|\Phi^{KS}[\rho]
\rangle \nonumber \\
 &=& \sum_{\sigma = \uparrow, \downarrow}
 \sum_{j=1}^{N_{\sigma}} \int \!
 \varphi_{j \sigma}^{KS}\left[\rho\right]({\bf r})^{\ast}
 \left(-\frac{1}{2} {\bf\nabla}^{2} \right)
 \varphi_{j\sigma}^{KS}\left[\rho\right]({\bf r})
 \, d^3 r \quad .
\eea
We mention in passing that the Hohenberg-Kohn theorem can also be
formulated for a ``Hartree-Fock world''~\cite{Pa79}, implying that the
HF density uniquely determines the external potential. Consequently
the HF ground-state determinant is a functional of the density as
well:
\be \label{PhiHF}
\Phi^{HF}[\rho] = \frac{1}{\sqrt{N!}} \det \left\{ \varphi_{j \sigma}^{HF}
[\rho] \right\} \quad .
\ee
The resulting kinetic-energy functional
\bea \label{THF}
T^{HF}[\rho] &=& \langle \Phi^{HF}[\rho]|\hat{T}|\Phi^{HF}[\rho]
\rangle \nonumber \\
 &=& \sum_{\sigma = \uparrow, \downarrow}
 \sum_{j=1}^{N_{\sigma}} \int \!
 \varphi_{j \sigma}^{HF}\left[\rho\right]({\bf r})^{\ast}
 \left(-\frac{1}{2} {\bf\nabla}^{2} \right)
 \varphi_{j\sigma}^{HF}\left[\rho\right]({\bf r})
 \, d^3 r
\eea
is different from $T_s[\rho]$ because the orbitals in (\ref{Ts}) come
from a {\em local} single-particle potential $v_s[\rho]$ while the
orbitals in (\ref{THF}) come from the {\em nonlocal} HF potential
$v^{HF}[\rho]$. However, the numerical difference between
$T^{HF}[\rho]$ and $T_s[\rho]$ has been found to be rather
small~\cite{GoEr95}.

The remaining term, $E_{xc}\left[ \rho \right]$, on the right hand side
of equation (\ref{1}) is termed the exchange-correlation (xc)
energy. Comparison of equation (\ref{1}) with equation (\ref{HK})
shows that the
xc-energy functional is formally given by
\be \label{Exc}
E_{xc}[\rho] =
\langle \Psi [\rho]|\,\hat{T}+\hat{W}_{Clb}\,|\Psi [\rho]\rangle
-T_s[\rho]
- \frac{1}{2} \int\!\int \! \frac{\rho({\bf r}) \rho({\bf r}')}
                           {|{\bf r - r'}|}     \, d^3r \, d^3r' \quad .
\ee

In density-functional theory the exact exchange-energy functional is
defined by
\be \label{7}
E_{x}^{DFT}[\rho] :=
\langle \Phi^{KS}[\rho]|\hat{W}_{Clb}|\Phi^{KS}[\rho]\rangle
- \frac{1}{2} \int\!\int \! \frac{\rho({\bf r}) \rho({\bf r}')}
                           {|{\bf r - r'}|}     \, d^3r \, d^3r' \quad .
\ee
This is identical with the ordinary Fock functional
\be \label{fock}
E_x^{HF} \left[ \varphi_{j \sigma} \right] = - \frac{1}{2}\sum_{\sigma
= \uparrow,
\downarrow} \sum_{j,k = 1}^{N_{\sigma}}
\int \! \int d^3 {r} \, d^3 {r'} \
 \frac{\varphi_{j\sigma}^{\ast}\br \varphi_{k\sigma}^{\ast}\brp
\varphi_{k\sigma}\br
\varphi_{j\sigma}\brp}{\vert {\bf r - r'} \vert}  \\
\ee
evaluated, however, with the KS Orbitals, i.~e.~
\be \label{Ex}
E_x^{DFT} \left[ \rho \right] =
E_x^{HF} \left[\varphi_{j \sigma}^{KS} \left[ \rho \right] \right] \quad .
\ee
The DFT correlation-energy functional is then given by
\be \label{8}
E_{c}^{DFT} = E_{xc}[\rho] - E_{x}^{DFT}[\rho] \quad .
\ee
Inserting the respective definitions (\ref{Exc}) and (\ref{Ex}) of
$E_{xc}[\rho]$ and $E_x^{DFT}[\rho]$ we find
\be \label{EcDFT}
E_{c}^{DFT}[\rho] = \langle \Psi [\rho]|\hat{T}+\hat{W}_{Clb}|\Psi [\rho]
\rangle -T_s[\rho]
- \frac{1}{2} \int\!\int \! \frac{\rho({\bf r}) \rho({\bf r}')}
                           {|{\bf r - r'}|}     \, d^3r \, d^3r'
-E_x^{HF} \left[\varphi_{j \sigma}^{KS} \left[ \rho \right] \right]
\,\, .
\ee
In terms of the Hartree-Fock total-energy functional
\begin{eqnarray} \label{ETOTHF}
E_{v_0}^{HF}\left[ \varphi_{j \sigma} \right] & = &
 \sum_{\sigma= \uparrow, \downarrow} \sum_{j=1}^{N \sigma} \int
\varphi_{j \sigma}({\bf r})^{\ast}
 \left(-\frac{1}{2} {\bf\nabla}^{2} \right) \varphi_{j \sigma}({\bf r})
 \,d^{3}r
 +  \int  \rho({\bf r}) \, v_0({\bf r}) \, d^3 {r} \nonumber \\
& & +  \frac{1}{2}  \int\! \int \frac{\rho({\bf r})
       \rho({\bf r'})}{\vert {\bf r - r' \vert}}
       \ d^3 {r} \ d^3 {r'}
 + E_x^{HF}\left[ \varphi_{j \sigma}\right]
\end{eqnarray}
and the total-energy functional (\ref{HK}) the DFT correlation energy
(\ref{EcDFT}) is readily expressed as
\be \label{19}
E_c^{DFT}[\rho] = E_{v_0}[\rho] -
E_{v_0}^{HF}\left[ \varphi_{j\sigma}^{KS} \left[ \rho \right] \right]
\quad .
\ee
By equation (\ref{EcexactDFT}), the exact DFT correlation energy is then
obtained by inserting the exact ground-state density $\rho_{exact}$
(corresponding to the external potential $v_0$)
into the functional (\ref{19}). By virtue of equation (\ref{Etot,exact})
one obtains
\be \label{20}
E_{c, exact}^{DFT} =
E_{tot, exact}
- E_{v_0}^{HF} \left[ \varphi_{j \sigma}^{KS}[\rho_{exact}] \right]
\quad .
\ee
The conventional quantum chemical correlation energy, on the other
hand, is given by
\be \label{21}
E_{c, exact}^{QC} =
E_{tot, exact} - E_{v_0}^{HF}
\left[\varphi_{j \sigma}^{HF} \left[ \rho_{HF} \right] \right]
\ee
where $\varphi_{j \sigma}^{HF}[\rho_{HF}]$ are the usual
selfconsistent HF orbitals corresponding to the external potential
$v_0$, i.~e.~ $\rho_{HF}$ is that very HF density which uniquely
corresponds to the external potential $v_0$. Of course, $\rho_{HF}$
and $\rho_{exact}$ are generally not identical.
Comparison of (\ref{20}) with (\ref{21}) shows that
\be \label{central}
E_{c, exact}^{DFT} = E_{c, exact}^{QC}
+ \left( E_{v_0}^{HF} \left[ \varphi_{j \sigma}^{HF} \left[ \rho_{HF}
\right] \right]
- E_{v_0}^{HF} \left[ \varphi_{j \sigma}^{KS}[\rho_{exact}] \right]
\right) \,.
\label{15}
\ee
This is the central equation relating the DFT correlation energy to
the QC correlation energy.
Since the HF orbitals $\varphi_{j \sigma}^{HF}[\rho_{HF}]$ are the
ones that minimize the HF total-energy functional (\ref{ETOTHF}), the
inequality
\begin{equation}
E_{v_0}^{HF} \left[ \varphi_{j \sigma}^{HF} \left[ \rho_{HF} \right] \right]
\leq E_{v_0}^{HF} \left[ \varphi_{j \sigma}^{KS}[\rho_{exact}] \right]
\end{equation}
must be satisfied and
it follows from equation (\ref{central}) that
\begin{equation} \label{15,5}
E_{c,exact}^{QC} \geq E_{c,exact}^{DFT} \quad .
\end{equation}
Equation (\ref{central}) tells us that, as a matter of principle,
selfconsistent DFT results for the
correlation energy should not be compared
directly with the conventional quantum chemical correlation energy but
rather with
the right-hand side of equation (\ref{15}). In
practise, of course,
quantum-chemical correlation energies and ground-state densities are known
only approximately, e.~g.~, from configuration-interaction (CI) calculations.
Hence,
\be \label{16}
E_{tot, CI} - E_{tot}^{HF} \left[ \varphi_{j \sigma}^{KS}[\rho_{CI}] \right]
\ee
is the quantity the selfconsistent DFT correlation energy should in
principle be compared
with. The second term of (\ref{16}) is readily computed by employing one of
the standard techniques~\cite{umri&gonz94,Go92,vanLee94,ZhMoPa94} of
calculating the KS potential
and its orbitals from a given CI density.
In the following we shall argue, however, that
the difference between $E_{c, exact}^{DFT}$ and $E_{c,
exact}^{QC}$ can be expected to be very small. To see this we rewrite
equation (\ref{15}) as
\begin{eqnarray}
E_{c, exact}^{DFT} -E_{c, exact}^{QC} &=&
\left( E_{v_0}^{HF} \left[ \varphi_{j \sigma}^{HF}\left[ \rho_{HF}
\right] \right]
- E_{v_0}^{HF} \left[ \varphi_{j \sigma}^{KS}[\rho_{x-only}] \right] \right)
\nonumber \\
&+& \left( E_{v_0}^{HF} \left[ \varphi_{j \sigma}^{KS}[\rho_{x-only}] \right]
- E_{v_0}^{HF} \left[ \varphi_{j \sigma}^{KS}[\rho_{exact}] \right]
\right) \,.
\label{17}
\end{eqnarray}
where $\rho_{x-only}$ is the ground-state density of an exact exchange-only
DFT calculation~\cite{enge&chev&macd&vosk92,enge&vosk93} and
$\varphi_{j \sigma}^{KS}[\rho_{x-only}]$ are the
corresponding KS orbitals.
The first difference on the right-hand side of equation (\ref{17}) is
known to be
small~\cite{enge&chev&macd&vosk92,enge&vosk93}. The second difference,
on the other hand, is
easily seen to be of {\em second} order in $(\rho_{x-only} -
\rho_{exact})$ and is therefore expected to be very small as well:
\begin{eqnarray}
\lefteqn{E_{v_0}^{HF} \left[ \varphi_{j \sigma}^{KS}[\rho_{x-only}] \right]
- E_{v_0}^{HF} \left[ \varphi_{j \sigma}^{KS}[\rho_{exact}] \right]}
\nonumber \\
& &= \int \! d^3r \left.
\frac{ \delta E_{v_0}^{HF} \left[\varphi_{j \sigma}^{KS}[\rho]\right] }
{ \delta \rho({\bf r}) } \right|_{\rho_{x-only}} \hspace{-3mm}
\cdot \left( \rho_{x-only}({\bf r}) - \rho_{exact}({\bf r}) \right)
+ O( \rho_{x-only} - \rho_{exact} )^{2} \nonumber \\
& &= \int \! d^3r \, \mu \cdot
\left( \rho_{x-only}({\bf r}) - \rho_{exact}({\bf r}) \right)
+ O( \rho_{x-only} - \rho_{exact} )^{2} \nonumber \\
& &= 0 + O( \rho_{x-only} - \rho_{exact} )^{2} \nonumber
\end{eqnarray}
The second equality follows from the fact that $\rho_{x-only}$ minimizes the
density functional $E_{v_0}^{HF} \left[\varphi_{j\sigma}^{KS}[\rho]\right]$.
Hence we conclude that $E_{c, exact}^{DFT} - E_{c, exact}^{QC}$ should
be small.
This estimate is confirmed by results of accurate variational
calculations on
H$^{-}$, He, Be$^{+2}$, Ne$^{+8}$~\cite{umri&gonz94},
as can be seen from Table 1.
There, the conventional quantum chemical correlation energies of
these systems are
compared with the ``exact'' DFT correlation energies calculated from equation
(\ref{20}).
For all elements and
ions shown, the difference between the two quantities is small, as
expected.
The values for ${\rm H^{-}}$, He and ${\rm Be^{+2}}$ also confirm the
relation (\ref{15,5}). For
Ne$^{+8}$,  where the difference is only \mbox{1 $\mu$H}, the inequality
(\ref{15,5}) is not satisfied, indicating inaccuracies of the
numerical procedure involved.

\begin{table}[hbtp]
\caption{ \sl Comparison of DFT correlation energies  with
 conventional quantum chemical correlation energies (QC).
$\Delta$ denotes the difference between the QC and the DFT correlation energy.
All numbers in Hartree units. }
\label{erste}
\begin{center}
\begin{tabular}{|l|l|l|l|}
\hline
& DFT & QC & \mbox{  $\Delta$} \\
\hline
${\rm H^{-}  }$& $-$0.041\,995 & $-$0.039\,821 & $+$0.002\,174 \\
He             & $-$0.042\,107 & $-$0.042\,044 & $+$0.000\,063 \\
${\rm Be^{+2}}$& $-$0.044\,274 & $-$0.044\,267 & $+$0.000\,007 \\
${\rm Ne^{+8}}$& $-$0.045\,692 & $-$0.045\,693 & $-$0.000\,001 \\
\hline
\end{tabular}
\end{center}
\end{table}
To conclude this section,  we mention that there exists yet another
possibility of defining a density functional for the correlation
energy~[4--11,13]:
\be \label{Ectilde}
\tilde{E}_c[\rho] =
E_{v_0}[\rho] - E_{v_0}^{HF}\left[ \varphi^{HF}_{j \sigma} \left[ \rho
\right] \right]
\ee
where $\varphi^{HF}_{j \sigma}[\rho]$ are the HF orbitals
corresponding to the density $\rho$ (see equation (\ref{PhiHF})).
If the exact density $\rho_{exact}$ is inserted in (\ref{Ectilde})
$\varphi^{HF}_{j \sigma}[\rho_{exact}]$ are the HF orbitals
corresponding to some unknown external potential $\tilde{v}_0$ whose
HF density is $\rho_{exact}$.
The decomposition
\be
\tilde{v}_0({\bf r}) =: v_0({\bf r}) + \tilde{v}_c({\bf r})
\ee
makes clear that on the single-particle level the definition
(\ref{Ectilde}) leads to a hybrid scheme featuring the ordinary {\em
non-local} HF exchange potential combined with the {\em local}
correlation potential $\tilde{v}_c({\bf r})$. In the present paper,
this hybrid scheme will not be further investigated.

\vspace{16pt}{\Large \bf Correlation Energies from Various DFT Approximations}

For further analysis, we compare in Tables \ref{ecsc0}, \ref{ecsc1} and
\ref{ecsc2} the DFT correlation energies resulting from various
approximations to $E_c^{DFT}[\rho]$.
LYP denotes the correlation-energy functional by Lee, Yang and
Parr~\cite{lee&yang&parr88}, PW91 the generalized gradient approximation by
Perdew and Wang~\cite{perd&wang92}, and LDA the conventional local
density approximation with the parametrisation of $E_{c}$ by Vosko,
Wilk and Nusair~\cite{vosk&wilk&nusa80}.
The first column,
denoted by OEP, shows the results of
 a recently developed scheme which employs an optimized effective
potential (OEP) including correlation effects
\cite{grab&gros95}. In this scheme
the full integral equation of the optimized effective potential method
\cite{shar&hort53, talm&shad76},
\begin{equation} \label{OEP-int}
\sum_{i=1}^{N_{\sigma}}  \int  d^{3} r' \left( V_{xc
\sigma}^{OEP}({\bf r'}) - u_{xc i \sigma}({\bf r'}) \right)
 \left( \sum_{\stackrel{k=1}{k \neq i}}^{\infty}
 \frac{\varphi_{k\sigma}^{\ast}({\bf r}) \varphi_{k\sigma}({\bf
 r'})}{\varepsilon_{k\sigma} - \varepsilon_{i\sigma} } \right)
 \varphi_{i\sigma}({\bf r}) \varphi_{i\sigma}^{\ast}( {\bf r'} )
 + c.c. = 0
\end{equation}
with
\begin{equation} \label{uxc}
u_{xc i \sigma}({\bf r})
:=
 \frac{ 1 }{ \varphi_{i\sigma}^{\ast} ({\bf r}) }
 \frac{ \delta E_{xc}\left[  \varphi_{j\sigma}  \right]
}{ \delta \varphi_{i\sigma}({\bf r}) }
\end{equation}
is solved semi-analytically by an approved method due to Krieger, Li and
Iafrate~\cite{krie&li&iafr92-1, krie&li&iafr92-2, krie&li&iafr93}:
\be \label{kli-eq}
 V_{xc\sigma}^{OEP}\br \approx
 V_{xc\sigma}^{KLI}\br =
 \frac{ 1 }{ \rho_{\sigma}({\bf r}) }
 \sum_{i=1}^{N_{\sigma}} \rho_{i\sigma}({\bf r})
 \left[ u_{xci\sigma}({\bf r}) + \left(\bar{V}_{xci\sigma}^{OEP}  -
        \bar{u}_{xci\sigma}  \right) \right]
\ee
where the constants $\left( \bar{V}_{xci\sigma}^{OEP} -
\bar{u}_{xci\sigma} \right)$ are the
solutions of the set of linear equations
\be \label{lin-eq}
\sum_{i=1}^{N_{\sigma}} \left( \delta_{ji} - M_{ji\sigma} \right) \left( \bar
V_{xci\sigma}^{OEP} - \bar u_{xci\sigma} \right)
=
\bar V_{xcj\sigma}^{S} - \bar u_{xcj\sigma}
\qquad j= 1, \ldots, N_{\sigma}
\ee
with
\be
M_{ji\sigma} := \int d^3 {r} \ \frac{\rho_{j\sigma}\br
\rho_{i\sigma}\br}{\rho_{\sigma}\br},
\ee
\be
V_{xc\sigma}^{S}\br := \sum_{i=1}^{N}
\frac{\rho_{i\sigma}\br}{\rho_{\sigma}\br} u_{xci\sigma}\br.
\ee
Here, $\bar u_{xcj\sigma}$ denotes the average value of
$u_{xcj\sigma}\br$ taken over the density of the $j\sigma$ orbital, i.~e.~
\be
\bar u_{xcj\sigma} = \int \rho_{j \sigma} \br
u_{xcj\sigma}\br d^3 {r}
\ee
and similarly for $\bar V_{xcj \sigma}^{S}$. Like in the conventional
Kohn-Sham method, the xc-potential resulting from equation
(\ref{kli-eq}) leads to a single-particle Schr\"odinger
equation with a {\em local} effective potential
\bea \label{1p-eq}
\left( -\frac{{\bf \nabla}^2}{2} +
v_0({\bf r}) +
 \int
 \frac{ \rho({\bf r'}) }{ \vert {\bf r} - {\bf r'} \vert }\,
 d^3 {r'} +
V_{xc \sigma}^{OEP}({\bf r})
\right) \varphi_{j\sigma} \br =
\varepsilon_{j \sigma} \varphi_{j \sigma} \br \\
{\textstyle (j = 1,\ldots,N_{\sigma} \quad \sigma = \uparrow,
\downarrow) .} \nonumber
\eea
The selfconsistent solutions $\varphi_{j\sigma}({\bf r})$
of equation (\ref{1p-eq})
with lowest single-particle energies $\varepsilon_{j \sigma}$
minimize the total-energy functional
\begin{eqnarray} \label{energie}
E_{v_0}^{OEP}\left[ \varphi_{j \sigma}  \right] & = &
 \sum_{\sigma = \uparrow, \downarrow} \sum_{i=1}^{N_{\sigma}} \int
\varphi_{i \sigma}^{\ast}({\bf r})
 \left(-\frac{1}{2} {\bf\nabla}^{2} \right) \varphi_{i \sigma}({\bf r})
 d^3 {r} \nonumber \\
& & +  \int \rho({\bf r}) \, v_0({\bf r})\, d^3{r} \nonumber \\
& & +  \frac{1}{2}  \int\! \int \frac{\rho({\bf r})
       \rho({\bf r'})}{\vert {\bf r - r' \vert}}
       \ d^3 {r} \, d^3 {r'} \nonumber \\
& & - \frac{1}{2}\sum_{\sigma = \uparrow,
\downarrow} \sum_{j,k = 1}^{N_{\sigma}}
\int \! \int d^3 {r}\, d^3 {r'} \
 \frac{\varphi_{j\sigma}^{\ast}\br \varphi_{k\sigma}^{\ast}\brp
\varphi_{k\sigma}\br
\varphi_{j\sigma}\brp}{\vert {\bf r - r'} \vert} \nonumber \\
& & + E_{c}^{CS}\left[ \{\varphi_{j \sigma} \} \right].
\end{eqnarray}
In the above equation, $E_{c}^{CS}$ denotes the Colle-Salvetti
functional~\cite{coll&sall75, coll&sall79} for the correlation-energy
given by
\bea \label{csec2}
E_{c}^{CS}  & = &
 - \ ab \int \gamma \br \xi \br \Biggl[ \sum_{\sigma}
\rho_{\sigma} \br \sum_{i} \mid \!{\bf \nabla} \varphi_{i\sigma} \br
\! \mid^{2}
\ - \ \ \frac{1}{4}  \mid \! {\bf \nabla} \rho \br \! \mid^{2}  \nonumber \\
& & - \ \frac{1}{4}  \sum_{\sigma}
\rho_{\sigma}\br \triangle \rho_{\sigma} \br \
+ \ \frac{1}{4}  \rho \br \triangle \rho \br \Biggr] d^3 {r} \nonumber \\
& & - \ a \int \gamma \br \frac{ \rho \br }{\eta \br} \ d^3 {r},
\eea
where
\begin{eqnarray}
\label{gam}
\gamma \br& = & 4 \ \frac{\rho_{\uparrow}\br
\rho_{\downarrow}\br}{\rho \br^{2}}, \\
\label{eta}
\eta \br & = & 1 + d \rho \br^{-\frac{1}{3}},\\
\label{xsi}
\xi \br & = & \frac{\rho \br^{-\frac{5}{3}} e^{-c \rho
\br^{-\frac{1}{3}}}}{\eta \br}.
\end{eqnarray}
The constants $a$, $b$, $c$ and $d$ are given by
\bed
\begin{array}{ll}
a = 0.04918, \qquad & b = 0.132, \\
c = 0.2533, & d = 0.349 .
\end{array}
\eed
\begin{table}[p]
\caption{ \sl Non-relativistic absolute correlation energies
resulting from various approximate DFT correlation energy functionals,
evaluated at the exact ground-state densities of the respective
atoms. ${|\Delta|}$ denotes the mean absolute deviation from the exact
DFT correlation energy. All numbers in Hartree units. }
\label{ecsc0}
\begin{center}
\begin{tabular}{|l|l|l|l|l|l|}
\hline
                   & OEP    &  LYP   &  PW91  & LDA    &  EXACT\\
\hline
     H$^{-}$       & 0.0297 & 0.0299 & 0.0320 & 0.0718 & 0.0420 \\
     He            & 0.0416 & 0.0438 & 0.0457 & 0.1128 & 0.0421 \\
     Be$^{+2}$     & 0.0442 & 0.0491 & 0.0537 & 0.1512 & 0.0443 \\
     Ne$^{+8}$     & 0.0406 & 0.0502 & 0.0617 & 0.2030 & 0.0457 \\
\hline
${|\Delta|}$
                   & 0.0045 & 0.0058 & 0.0097 & 0.0912 &        \\
\hline
\end{tabular}
\end{center}
\end{table}

\begin{table}[p]
\caption{ \sl Non-relativistic absolute
correlation energies of first and second row atoms
from selfconsistent calculations with various DFT approximations. QC
denotes the conventional quantum chemistry value.
${|\Delta|} \%$ denotes the mean value of $\left|
(E_c^{DFT}-E_c^{QC})/{E_c^{QC}} \right|$ in percent.
All other numbers in Hartree units. }
\label{ecsc1}
\begin{center}
\begin{tabular}{|l|r|r|r|r||r|}
\hline
 & OEP & LYP & PW91 & LDA & \mbox{QC  } \\
\hline
     He         &     0.0416 &     0.0437 &     0.0450 & 0.1115 & 0.0420 \\
     Li         &     0.0509 &     0.0541 &     0.0571 & 0.1508 & 0.0453 \\
     Be         &     0.0934 &     0.0954 &     0.0942 & 0.2244 & 0.0943 \\
     B          &     0.1289 &     0.1287 &     0.1270 & 0.2906 & 0.1249 \\
     C          &     0.1608 &     0.1614 &     0.1614 & 0.3587 & 0.1564 \\
     N          &     0.1879 &     0.1925 &     0.1968 & 0.4280 & 0.1883 \\
     O          &     0.2605 &     0.2640 &     0.2587 & 0.5363 & 0.2579 \\
     F          &     0.3218 &     0.3256 &     0.3193 & 0.6409 & 0.3245 \\
     Ne         &     0.3757 &     0.3831 &     0.3784 & 0.7434 & 0.3905 \\
     Na         &     0.4005 &     0.4097 &     0.4040 & 0.8041 & 0.3956 \\
     Mg         &     0.4523 &     0.4611 &     0.4486 & 0.8914 & 0.4383 \\
     Al         &     0.4905 &     0.4979 &     0.4891 & 0.9661 & 0.4696 \\
     Si         &     0.5265 &     0.5334 &     0.5322 & 1.0418 & 0.5050 \\
     P          &     0.5594 &     0.5676 &     0.5762 & 1.1181 & 0.5403 \\
     S          &     0.6287 &     0.6358 &     0.6413 & 1.2259 & 0.6048 \\
     Cl         &     0.6890 &     0.6955 &     0.7055 & 1.3289 & 0.6660 \\
     Ar         &     0.7435 &     0.7515 &     0.7687 & 1.4296 & 0.7223 \\
\hline
${|\Delta|} \%$
	        &     3.13   &     4.52   &     5.10   & 120    &   \\
\hline
\end{tabular}
\end{center}
\end{table}

\begin{table}[htb]
\caption{ \sl Non-relativistic absolute correlation energies of atoms
from selfconsistent calculations with various DFT approximations.
All numbers in Hartree units. }
\label{ecsc2}
\begin{center}
\begin{tabular}{|l|r|r|r||l|r|r|r|}
\hline
 & OEP & LYP & PW91 &  & OEP & LYP & PW91  \\
\hline
     K   &0.8030 &0.7821 &0.7994  & Rb  &1.7688 &1.7832 &1.9509  \\
     Ca  &0.8269 &0.8329 &0.8467  & Sr  &1.8222 &1.8355 &2.0056  \\
     Sc  &0.8832 &0.8855 &0.9033  & Y   &1.8763 &1.8863 &2.0671  \\
     Ti  &0.9371 &0.9374 &0.9613  & Zr  &1.9281 &1.9363 &2.1307  \\
     V   &0.9882 &0.9882 &1.0198  & Nb  &1.9475 &1.9558 &2.1899  \\
     Cr  &1.0073 &1.0086 &1.0736  & Mo  &1.9905 &2.0003 &2.2551  \\
     Mn  &1.0812 &1.0861 &1.1375  & Tc  &2.0796 &2.0874 &2.3412  \\
     Fe  &1.1597 &1.1620 &1.2158  & Ru  &2.1571 &2.1637 &2.4254  \\
     Co  &1.2324 &1.2331 &1.2933  & Rh  &2.2278 &2.2340 &2.5081  \\
     Ni  &1.3009 &1.3010 &1.3700  & Pd  &2.3123 &2.3154 &2.6074  \\
     Cu  &1.3693 &1.3694 &1.4562  & Ag  &2.3561 &2.3649 &2.6705  \\
     Zn  &1.4273 &1.4303 &1.5212  & Cd  &2.4146 &2.4247 &2.7373  \\
     Ga  &1.4704 &1.4753 &1.5768  & In  &2.4600 &2.4704 &2.7964  \\
     Ge  &1.5101 &1.5174 &1.6343  & Sn  &2.5024 &2.5135 &2.8577  \\
     As  &1.5465 &1.5570 &1.6917  & Sb  &2.5419 &2.5544 &2.9193  \\
     Se  &1.6177 &1.6288 &1.7662  & Te  &2.6134 &2.6252 &2.9965  \\
     Br  &1.6795 &1.6912 &1.8393  & I   &2.6763 &2.6876 &3.0726  \\
     Kr  &1.7355 &1.7493 &1.9112  & Xe  &2.7338 &2.7456 &3.1475  \\
\hline
\end{tabular}
\end{center}
\end{table}

\vspace*{-2mm}
In Table \ref{ecsc0}, the four {\em approximate} DFT
correlation energy functionals are evaluated at the exact
densities~\cite{umri&gonz94} of
H$^{-}$, He, Be$^{+2}$, Ne$^{+8}$ and compared
with the {\em exact} DFT correlation energies given by equation (\ref{20}).
On average, the OEP values are superior.

In Table \ref{ecsc1} selfconsistent DFT correlation energies are compared
with QC values taken from~\cite{chak&gwal&davi&parp93}.
In these selfconsistent calculations the {\em approximate}
correlation-energy functionals $E_c^{LYP}$, $E_c^{PW91}$, $E_c^{LDA}$
are complemented with the {\em approximate} exchange-energy functionals
$E_x^{B88}$~\cite{beck88}, $E_x^{PW91}$~\cite{perd&wang92} and
$E_x^{LDA}$, respectively. In the OEP
case, the DFT exchange-energy functional (\ref{Ex}) is of course
treated exactly. The numerical data show three main features.
\begin{enumerate}

\item For most atoms, the
absolute value of $E_{c}^{QC}$ is
smaller than the absolute correlation energy obtained with any DFT
method, as it should be according to the relation (\ref{15,5}).

\item The values of $E_c^{OEP}$, $E_c^{LYP}$, $E_c^{PW91}$ and
$E_c^{QC}$ agree quite closely with each other while
the absolute value of $E_{c}^{LDA}$ is
too large roughly by a factor of two.
We mention that due to the well known
error cancellation between $E_{x}^{LDA}$ and
$E_{c}^{LDA}$, the resulting LDA values for total xc energies are
much better.

\item The difference between $E_{c}^{DFT}$ and $E_{c}^{QC}$ is
smallest for the $E_c^{OEP}$ values, larger for $E_c^{LYP}$
and largest for $E_c^{PW91}$.
The difference between $E_c^{QC}$ and $E_c^{DFT}$ has three
sources:
\begin{enumerate}
\item The values of $E_c^{QC}$ are only approximate, i.~e.~not
identical with $E_{c, exact}^{QC}$.
\item The values of $E_c^{DFT}$ are only approximate, i.~e.~not
identical with $E_{c, exact}^{DFT}$.
\item As shown in the last section, the exact values $E_{c,
exact}^{QC}$ and $E_{c, exact}^{DFT}$ are not identical.
\end{enumerate}
Currently it is not known with certainty which effect gives the
largest contribution. However, with the arguments given in the last
section, we
expect the contribution of (c) to be small. Assuming that the quoted
values of $E_{c}^{QC}$ are very close to $E_{c, exact}^{QC}$ we
conclude that $E_c^{OEP}$ is closest to $E_{c, exact}^{DFT}$.

\end{enumerate}

Table \ref{ecsc2} shows correlation energies of atoms K through Xe
obtained with the various selfconsistent DFT approaches.
In almost all
cases, the absolute OEP values for $E_{c}$ are smallest and the ones from
PW91 are largest, while the LYP values lie in between.
In most cases, $E_c^{OEP}$ and $E_c^{LYP}$ agree within less than 1 \%
while $| E_c^{PW91} |$ is larger (by up to 10 \%) as the atomic number
$Z$ increases.
We emphasize that reliable values
for $E_{c}^{QC}$ do not exist for these atoms.
\newpage

\vspace{16pt}{\Large \bf Acknowledgments}

We thank C.~Umrigar for providing us with the exact densities and KS
potentials for H$^{-}$, He, Be$^{+2}$ and Ne$^{+8}$.
We gratefully appreciate the help of Dr.~E.~Engel especially for
providing us with a Kohn-Sham computer code and
for some helpful discussions. We
would also like to thank Professor J.~Perdew for providing us with
the PW91 xc subroutine.
This work
was supported in part by the Deutsche Forschungsgemeinschaft.


\end{document}